\documentclass[12pt]{iopart}
\usepackage{enumerate}
\usepackage{subfigure}
\usepackage{iopams}
\usepackage{amstext}
\usepackage{psfrag}
\usepackage{graphicx}

\begin{document}

\title{Current fluctuations in the zero-range process with open boundaries}
\author{R J Harris, A R{\'a}kos\footnote{Present address: Department of Physics of Complex Systems, Weizmann Institute of Science, Rehovot, Israel 76100.} and G M Sch{\"u}tz}
\address{Institut f\"ur Festk\"orperforschung, Forschungszentrum J\"ulich, 52425 J\"ulich, Germany}

\ead{r.harris@fz-juelich.de}

\begin{abstract}

We discuss the long-time limit of the integrated current distribution for the one-dimensional zero-range process with open boundaries.  We observe that the current fluctuations become site-dependent above some critical current and argue that this is a precursor of the condensation transition which occurs in such models.  Our considerations for the totally asymmetric zero-range process are complemented by a Bethe ansatz treatment for the equivalent exclusion process.

\end{abstract}


\submitto{Journal of Statistical Mechanics: Theory and Experiment}

\maketitle

\section{Introduction}

The zero-range process (ZRP) is a simple lattice-based model in which particles move randomly to neighbouring lattice sites with rates depending only on the occupation of the departure site.  This model was first introduced by Spitzer in 1970~\cite{Spitzer70}; more recent interest results from the observation that, for particular choices of hopping parameters the model exhibits a condensation phenomenon~\cite{Evans96,Krug96,OLoan98,Evans00,Jeon00} where a macroscopic proportion of particles accumulate on a single site (a real-space analogue of Bose-Einstein condensation).  Condensation transitions occur in a wide variety of physical and non-physical contexts~\cite{Shim04,Burda02,Frohlich75,Chowdhury00,Helbing01,Bianconi01,Dorogovtsev03c} and analytical studies of ``toy'' models, such as the ZRP, can provide important insight. 
The one-dimensional (1D) zero-range process has additional significance since it can be mapped to well-studied exclusion processes~\cite{Mukamel00,Schutz03}.  Condensation in the ZRP then corresponds to phase separation in the exclusion process~\cite{Kafri02c}.
For detailed discussion of recent progress on the ZRP and related models see, e.g., the review by Evans and Hanney~\cite{Evans05}.

Most of these previous works study the ZRP defined on a 1D periodic or infinite lattice.  In a recent paper Levine~\emph{et al.}~\cite{Levine04c} considered the case with open boundary conditions, deriving the steady-state distribution and characterizing the condensation which occurs for strong boundary drive.  Our contribution is to study the large-time limit of the current fluctuations through such an open 1D ZRP with fixed initial configuration.  This is in the spirit of previous studies of current fluctuations in the exclusion process~\cite{Derrida04b} and more general boundary-driven lattice gases~\cite{Wijland04}.

Specifically, the aim is to give the integrated current distribution and hence derive the large deviation properties of the measured time-averaged current.  This helps to give insight into the nature of non-equilibrium steady states where large deviation functionals play a r\^ole similar to the free energy of equilibrium systems~\cite{Derrida01}.  For example, there has been much recent interest in ``fluctuation theorems''~\cite{Gallavotti95, Kurchan98, Lebowitz99, Gaspard04,Seifert04}.  By considering the dynamics under time-reversal, such theorems characterize the possible fluctuations of driven systems beyond the linear response regime and impose a symmetry property (to be discussed below) on the large deviation function.  Essentially, this symmetry relates the probability of observing a given rate of entropy increase to the probability of observing the same rate of entropy decrease, leading to verifiable predictions for simulation~\cite{Evans93} and experiment~\cite{Ciliberto04}. 

For the zero-range model considered here, our main physical finding is that, even for parameter values which result in a well-defined stationary state (i.e., no condensation), the current fluctuations undergo a change in character above some critical current.  We interpret this qualitatively as a result of the temporary build-up of particles on some site(s) and argue that it is a precursor of the condensation transition.  For unidirectional particle-hopping we show that the current distributions become spatially inhomogeneous for high currents.  In this totally asymmetric case we also explicitly demonstrate the link to current fluctuations in the totally asymmetric simple exclusion process which we solve via the Bethe ansatz with particle-dependent hopping rates~\cite{Rakos05}.  

The remainder of the paper is structured as follows.  In section~\ref{s:ZRP} we define the zero-range model and summarize known steady-state results~\cite{Levine04c} within the framework of the quantum Hamiltonian formalism.  Section~\ref{s:cf} contains a calculation of the integrated current distribution for the general ZRP (arbitrary hopping rates and boundary parameters) together with a discussion of the regime of its validity.   In section~\ref{s:TAZRP} we gain deeper understanding by considering in detail the behaviour of current fluctuations in the totally asymmetric zero-range process.  Then, in section~\ref{s:TASEP}, we present a determinant solution for the current distribution in this totally asymmetric model, obtained by utilizing Bethe ansatz results for the equivalent exclusion process.   Finally, in section~\ref{s:dis} we summarize our work and discuss some related issues.

\section{Zero-range process with open boundaries}
\label{s:ZRP}

\subsection{Definition of model}
\label{ss:df}

\begin{figure}
\begin{center}
\psfrag{1}[][]{1}
\psfrag{2}[][]{2}
\psfrag{3}[][]{3}
\psfrag{4}[][]{4}
\psfrag{5}[][]{5}
\psfrag{L-2}[][]{$L\!-\!2$}
\psfrag{L-1}[][]{$L\!-\!1$}
\psfrag{L}[][]{$L$}
\psfrag{a}[][]{$\alpha$}
\psfrag{d}[][]{$\delta$}
\psfrag{bw}[][]{$\beta w_n$}
\psfrag{cw}[][]{$\gamma w_n$}
\psfrag{pw}[][]{$p w_n$}
\psfrag{qw}[][]{$q w_n$}
\includegraphics*[width=0.9\textwidth]{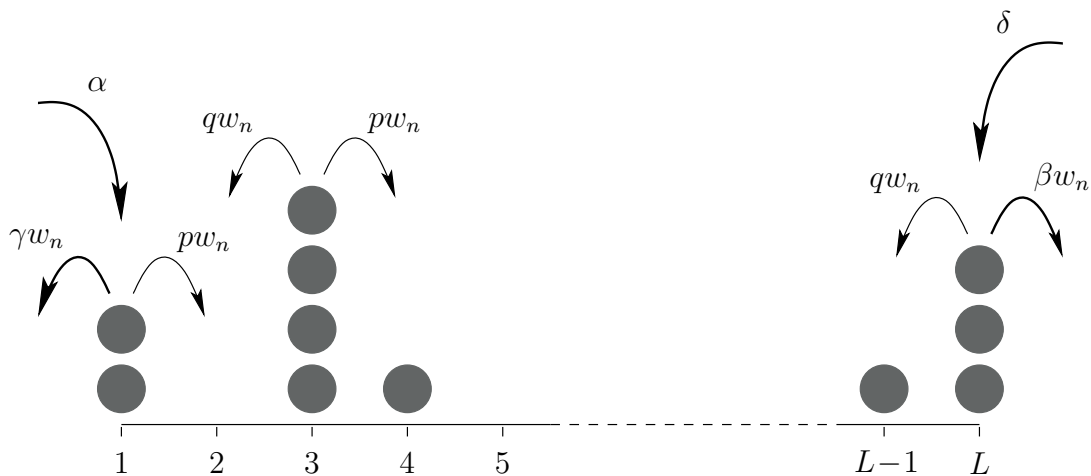}
\caption{Schematic representation of the ZRP on an open $L$-site lattice}
\label{f:ZRP}
\end{center}
\end{figure}

We consider the ZRP defined on an 1D open lattice of $L$ sites---illustrated in figure~\ref{f:ZRP}.
Each site $l$ contains an arbitrary integer number of particles $n$ which hop randomly to neighbouring sites (with exponentially distributed waiting time).   In the bulk the topmost particle from site $l$ moves to the right (left) with rate $p w_n$ ($q w_n$) where $w_n$ depends only on the occupation number $n$ of the departure site and, by definition, $w_0 = 0$.  Particles are injected onto site 1 ($L$) with rate $\alpha$ ($\delta$) and removed with rate $\gamma w_n$ ($\beta w_n$).  To facilitate later discussion of current fluctuations across different bonds we will label each bond by the site at its lefthand end, i.e., the $l$th bond is between sites $l$ and $l+1$.

For particular choices of $w_n$, the model with periodic boundary conditions exhibits a condensation transition (see, e.g.,~\cite{OLoan98, Evans00, Kafri02c,Jeon00b,Grosskinsky03,Godreche03}) where, above some critical density $\rho_c$, excess particles all accumulate on one randomly chosen site while the remaining sites have average density $\rho_c$.  In~\cite{Levine04c} it is demonstrated that (even for choices of $w_n$ which do not give condensation on a ring geometry) open boundary conditions with strong driving can result in condensation at one or both of the boundary sites.  In the present work, we first discuss current fluctuations for open boundary conditions with arbitrary $w_n$ (section~\ref{s:cf}) before specializing to the totally asymmetric case with $w_n=1$ (section~\ref{s:TAZRP}).  This latter model can be mapped to the totally asymmetric exclusion process as discussed in section~\ref{s:TASEP}.

The zero-range process can be conveniently represented using the quantum Hamiltonian formalism~\cite{Schutz01}.  In this approach one works in the vector space $(\mathbb{C}^\infty)^{\otimes L}$, defining a probability vector $|P\rangle = \sum_n P_n |n\rangle$ with $|n\rangle$ the basis vector associated with the particle configuration $n=(n_1,n_2,\ldots,n_L)$ and $P_n$ the probability measure on the set of all such configurations. $|P\rangle$ obeys the normalization condition $\langle s|P \rangle =1$ where $\langle s | = \sum_n \langle n |$ and $\langle n | n' \rangle = \delta_{n,n'}$.  Within this formalism the ZRP time evolution is represented by the Master equation
\begin{equation}
\frac{\rmd}{\rmd t}|P(t)\rangle = - H |P(t)\rangle
\end{equation}
with
\begin{eqnarray}
\fl H=- \biggl\{ \sum_{l=1}^{L-1} \left[  p(a_l^- a^+_{l+1} -d_l) + q (a_l^+ a_{l+1}^- - d_{l+1})\right] \nonumber \\ 
+ \alpha(a_1^+-1) + \gamma(a_1^- - d_1)  +\delta (a_L^+-1)+\beta(a^-_L-d_L) \biggr\} \label{e:H}
\end{eqnarray}
where $a^+$ and $a^-$ are infinite-dimensional particle creation and annihilation matrices
\begin{equation}
\fl a^+=\left(\begin{array}{ccccc}
0 & 0 & 0 & 0 & \ldots \\
1 & 0 & 0 & 0 & \ldots \\
0 & 1 & 0 & 0 & \ldots \\
0 & 0 & 1 & 0 & \ldots \\
\ldots & \ldots & \ldots & \ldots & \ldots
\end{array}
\right),
\qquad
a^-=\left(\begin{array}{ccccc}
0 & w_1 & 0 & 0 & \ldots \\
0 & 0 & w_2 & 0 & \ldots \\
0 & 0 & 0 & w_3 & \ldots \\
0 & 0 & 0 & 0 & \ldots \\
\ldots & \ldots & \ldots & \ldots & \ldots
\end{array} \label{e:aa+}
\right)
\end{equation}
and $d$ is a diagonal matrix with the $(i,j)$th element given by $w_i \delta_{i,j}$.

\subsection{Stationary state}
\label{ss:ss}

The steady state of a stochastic process corresponds to the ground state of the quantum Hamiltonian.  The left eigenvector with zero eigenvalue is just $\langle s |$, i.e., the constant row vector $(1,1,1,\ldots 1)$.  The corresponding right eigenvector obeys
\begin{equation}
H|P^*\rangle=0.
\end{equation}
In~\cite{Levine04c} it is shown that, for the ZRP, the stationary distribution is given by a product measure
\begin{equation}
|P^*\rangle = |P^*_1) \otimes |P^*_2) \otimes \ldots \otimes |P^*_L) \label{e:prod}
\end{equation}
where $|P^*_l)$ is the probability vector with components
\begin{equation}
P^*(n_l=n) = \frac{z_l^n}{Z_l}\prod_{i=1}^n w_i^{-1} \label{e:prob}
\end{equation}
with the empty product $n=0$ defined equal to 1.  Here $Z_l$ is the local analogue of the grand-canonical partition function
\begin{equation}
Z_l \equiv Z(z_l)=\sum_{n=0}^{\infty} z_l^n \prod_{i=1}^{n} w_i^{-1} \label{e:gc}
\end{equation}
and the fugacities $z_l$ are uniquely defined by the following stationarity condition
\begin{equation}
pz_l-qz_{l+1}=\alpha-\gamma z_1 = \beta z_L - \delta \equiv c \label{e:zss}
\end{equation}
with solution~\cite{Levine04c}
\begin{equation}
z_l=\frac{[(\alpha+\delta)(p-q)-\alpha\beta+\gamma\delta]\left(\frac{p}{q}\right)^{l-1} - \gamma\delta + \alpha\beta \left(\frac{p}{q}\right)^{L-1}}{\gamma(p-q-\beta) + \beta(p-q+\gamma)\left(\frac{p}{q}\right)^{L-1}} \label{e:zss2}
\end{equation}
and mean steady-state current
\begin{equation}
c=(p-q)\frac{- \gamma\delta + \alpha\beta \left(\frac{p}{q}\right)^{L-1}}{\gamma(p-q-\beta) + \beta(p-q+\gamma)\left(\frac{p}{q}\right)^{L-1}}. \label{e:css}
\end{equation}
These results will be important for the calculations of the following sections.  Note that the existence of the steady state, for given $w_n$, is determined by the radius of convergence of $Z$.  For parameters resulting in some $z_l$ outside this radius of convergence, a growing condensate occurs~\cite{Levine04c}.

\section{Current fluctuations}
\label{s:cf}

\subsection{Definitions and notation}
\label{ss:cfintro}

We are interested in the probability distribution of the total integrated current $J_l(t)$, i.e., the net number of particle jumps between sites $l$ and $l+1$ in the time interval $[0,t]$.   It turns out to be convenient to study the generating function $\langle \rme^{-\lambda J_l(t)} \rangle$ where the angled brackets denote an average over histories.  In particular, we wish to find an explicit expression in terms of system parameters for the quantity $e_l(\lambda)$ defined by 
\begin{equation}
e_l(\lambda) =\lim_{t \rightarrow \infty} - \frac{1}{t} \ln {\langle \rme^{-\lambda J_l(t)} \rangle}. \label{e:e_l}
\end{equation}
This characterizes the asymptotic current distribution which, for an ergodic system, is not expected to depend on the choice of initial particle configuration.

As observed in~\cite{Lebowitz99}, equation~(\ref{e:e_l}) implies a large deviation property for the probability distribution, $p_l(j,t)=\mathrm{Prob}(j_l=j,t)$, of the observed ``average'' current $j_l=J_l/t$.  The long-time limiting behaviour is given by
\begin{equation}
p_l(j,t) \sim \rme^{-t\hat{e}_l(j)} \label{e:pj}
\end{equation}
where $\hat{e}_l(j)$ is the Legendre transformation of $e_l(\lambda)$ \footnote{This can be shown by performing a Fourier transform on the generating function and then using a saddle-point approximation to evaluate the resulting integral for large times.}
\begin{equation}
\hat{e}_l(j)=\max_{\lambda}\{e_l(\lambda)-\lambda j \}. \label{e:lang}
\end{equation}

We use the notation $e_l(\lambda)$, $\hat{e}_l(j)$ and $p_l(j,t)$ to indicate the spatial dependence of the current distribution.  It is clear that for a well-defined steady state the mean current must be spatially constant, i.e, $\langle {j}_l \rangle = c$ for all $l$, but it does not necessarily follow that the full distribution of asymptotic current fluctuations is translationally invariant.  We will return to this point later. 

\subsection{Integrated distribution of input current}
\label{ss:cfin}

Let us start by considering current fluctuations into the system, i.e., across the bond between the ``reservoir site'' 0 and the first proper site 1.  We represent the total integrated current across the ``0th'' bond by the diagonal matrix $J_0$ (whose diagonal elements are the set of all integer values) which acts on a state with basis vectors corresponding to the actual value of the integrated current across this bond.  Then we introduce raising and lowering operators $X^\pm$ on this space.  These obey the commutation relation $[X^\pm, J_0] =\mp X^\pm$ which implies
\begin{equation}
\rme^{-\lambda J_0} X^{\pm} \rme^{\lambda J_0} = \rme^{\mp \lambda} X^{\pm}. \label{e:Xcomm}
\end{equation}
In the joint configuration/current state space the Hamiltonian is given by
\begin{eqnarray}
\fl H_0=- \biggl\{ \sum_{l=1}^{L-1} \left[  p(a_l^- a^+_{l+1} -d_l) + q (a_l^+ a_{l+1}^- - d_{l+1})\right] \nonumber \\
+ \alpha(X^+ a_1^+-1) + \gamma(X^- a_1^- - d_1)  +\delta (a_L^+-1)+\beta(a^-_L-d_L) \biggr\}.
\end{eqnarray}

Now we wish to measure the expectation value
\begin{equation}
\langle \rme^{-\lambda J_0(t)} \rangle = \langle s,s| \rme^{-\lambda J_0} \rme^{-H_0 t} |P,0 \rangle
\end{equation}
where the extra state labels refer to the current space and $|P\rangle$ is an arbitrary specific initial particle configuration obeying the normalization condition $\langle s | P \rangle = 1$.  Using the transformation~(\ref{e:Xcomm}) one obtains 
\begin{equation}
\langle \rme^{-\lambda J_0(t)} \rangle = \langle s | \rme^{-\tilde{H}_0 t} |P\rangle
\end{equation}
with
\begin{eqnarray}
\fl \tilde{H}_0=- \biggl\{ \sum_{l=1}^{L-1} \left[  p(a_l^- a^+_{l+1} -d_l) + q (a_l^+ a_{l+1}^- - d_{l+1})\right]  \nonumber \\
+ \alpha(a_1^+ \rme^{-\lambda}-1) + \gamma(a_1^- \rme^{\lambda} - d_1)  +\delta (a_L^+-1)+\beta(a^-_L-d_L) \biggr\}. \label{e:H0}
\end{eqnarray}
In other words, the terms in the original Master equation which give a unit increase/decrease in the integrated input current are simply multiplied by  $e^{\mp\lambda}$.  We are particularly concerned with the long-time limiting behaviour which is given by
\begin{equation}
\lim_{t \rightarrow \infty} \langle \rme^{-\lambda J_0(t)} \rangle = \langle s |0 \rangle \langle 0 | P\rangle \rme^{-\tilde{e}_0(\lambda) t} \label{e:pre}
\end{equation}
where $\tilde{e}_0(\lambda)$ is the lowest eigenvalue of the modified non-stochastic Hamiltonian~(\ref{e:H0}) and $|0\rangle$ is the corresponding ground-state eigenvector.  If the time-independent prefactor on the righthand side of equation~(\ref{e:pre}) is finite (meaning that the ground state is normalizable in the sense that $\langle 0 |0 \rangle$ and $\langle s |0 \rangle$ are both finite) then we can identify $\tilde{e}_0(\lambda)$ with $e_0(\lambda)$, as defined by equation~(\ref{e:e_l}).  

To determine $e_0(\lambda)$ for this case we make the ansatz that the ground state is still a product state of the form~(\ref{e:prod})--(\ref{e:gc}) but with modified fugacities and then use the following properties of creation and annihilation operators
\begin{eqnarray}
a_l^+ |0\rangle=z_l^{-1} d_l |0\rangle \\
a_l^- |0\rangle=z_l |0\rangle
\end{eqnarray}
which can be explicitly checked with the matrices given in~(\ref{e:aa+}).  Then $|0\rangle$ satisfies
\begin{eqnarray}
\fl -\tilde{H}_0|0\rangle = \biggl[ \sum_{l=1}^{L-1} (p z_l - q z_{l+1})(z_{l+1}^{-1} d_{l+1} - z_l^{-1} d_l) \nonumber \\
+ (\alpha \rme^{-\lambda} - \gamma z_1)z_1^{-1}d_1 + (\delta - \beta z_L) z_L^{-1}d_L - \alpha+\gamma \rme^{\lambda}z_1 - \delta + \beta z_L \biggr] |0\rangle. 
\end{eqnarray}
It is clear that if $|0\rangle$ is an eigenstate the coefficients of the $d$'s must cancel, leading to the condition
\begin{equation}
pz_l-qz_{l+1}=\alpha \rme^{-\lambda} -\gamma z_1 = \beta z_L - \delta.
\end{equation}
Comparison with equation~(\ref{e:zss}) shows that the required eigenstate of $\tilde{H}_0$ is just given by the stationary product state of the ZRP with the replacement $\alpha \rightarrow \alpha \rme^{-\lambda}$, i.e., the fugacities are 
\begin{equation}
z_l=\frac{[(\alpha \rme^{-\lambda} +\delta)(p-q)-\alpha\beta \rme^{-\lambda} +\gamma\delta]\left(\frac{p}{q}\right)^{l-1} - \gamma\delta + \alpha\beta \rme^{-\lambda} \left(\frac{p}{q}\right)^{L-1}}{\gamma(p-q-\beta) + \beta(p-q+\gamma)\left(\frac{p}{q}\right)^{L-1}}. \label{e:fug}
\end{equation}
The eigenvalue of this state is given by
\begin{equation}
e_0(\lambda) = - \alpha+\gamma \rme^{\lambda}z_1 - \delta + \beta z_L
\end{equation}
and using~(\ref{e:fug}) then yields
\begin{equation}
e_0(\lambda) =\frac{(p-q)(\rme^\lambda-1)\left[\alpha\beta\left(\frac{p}{q}\right)^{L-1}\rme^{-\lambda}-\gamma\delta\right]}{\gamma(p-q-\beta)+\beta(p-q+\gamma)\left(\frac{p}{q}\right)^{L-1}}. \label{e:elamb}
\end{equation}

Since $\tilde{H}_0$ is not stochastic the left ground state $\langle 0 |$ is not necessarily $\langle s|$.  In fact, using analogous methods to those employed above we can show that, for any choice of $w_n$, it is a product state with $(0_l| = (1, \tilde{z}_l, {\tilde{z}_l}^2, \dots)$ where 
\begin{equation}
\tilde{z}_l=\frac{\beta\gamma(\rme^\lambda-1)\left(\frac{p}{q}\right)^{L-l} + \gamma \rme^\lambda (p-q-\beta) + \beta (p-q+\gamma)\left(\frac{p}{q}\right)^{L-1} }{\gamma(p-q-\beta) + \beta(p-q+\gamma)\left(\frac{p}{q}\right)^{L-1}}. \label{e:lfug}
\end{equation}

As $\lambda \rightarrow 0$, we see that $|0\rangle \rightarrow |P^*\rangle$, $\langle 0 | \rightarrow \langle s|$ and $e_0(\lambda) \rightarrow 0$, as expected.  To argue that this product measure is indeed the required lowest eigenstate of $\tilde{H}_0$ for all $\lambda$ we appeal to Perron-Frobenius theory~\cite{Frobenius12,Minc88}.  
We note that $\tilde{H}_0$ contains positive diagonal elements together with non-positive off-diagonal elements. It follows that the time evolution operator $\rme^{-\tilde{H}_0 t}$ is a non-negative irreducible matrix which must have a real maximal eigenvalue corresponding to its only non-negative eigenvector.  
Now, it is straightforward to show that for all parameter values $z_l>0$ (and $\tilde{z}_l>0$), so we have here found the positive eigenvector and~(\ref{e:elamb}) must therefore be the lowest eigenvalue of $\tilde{H}_0$.  However, note that this argument is not mathematically rigorous for an infinite-dimensional matrix.

Equation~(\ref{e:elamb}) for $e_0(\lambda)$ is the chief result of this section and after Legendre transformation we obtain the rather cumbersome explicit expression for $\hat{e}_0(j)$:
\begin{eqnarray}
\fl \hat{e}_0(j) =  \frac{(p-q)[\alpha \beta (p/q)^{L-1}+\gamma\delta]}{\gamma(p-q-\beta)+\beta(p-q+\gamma)(p/q)^{L-1}} \nonumber \\
-\sqrt{j^2 + \frac{4\alpha\beta\gamma\delta(p/q)^{L-1}(p-q)^2}{[\gamma(p-q-\beta)+\beta(p-q+\gamma)(p/q)^{L-1}]^2}} \nonumber \\
- j \ln \left[ \frac{2\alpha\beta (p/q)^{L-1} (p-q)}{\gamma(p-q-\beta)+\beta(p-q+\gamma)(p/q)^{L-1}} \right] \nonumber \\
+ j \ln \left[ j+ \sqrt{j^2 + \frac{4\alpha\beta\gamma\delta(p/q)^{L-1}(p-q)^2}{[\gamma(p-q-\beta)+\beta(p-q+\gamma)(p/q)^{L-1}]^2}}\right]. 
\end{eqnarray}
In the following subsection we further discuss the regime of validity of~(\ref{e:elamb}), but first we make some comments on the form of our result.

For the bulk symmetric case with driving boundary conditions taking the limit $p=q=1$ in~(\ref{e:elamb}) gives
\begin{equation}
e_0(\lambda) = \frac{(\rme^\lambda - 1)(\rme^{-\lambda} \alpha \beta - \gamma \delta)}{\gamma + \beta + \beta \gamma (L-1)}.
\end{equation}
Van Wijland and Racz~\cite{Wijland04} previously calculated the leading order term in the $L\rightarrow \infty$ expansion of this result by field theoretic methods. It has also been obtained by Bodineau and Derrida~\cite{Bodineau04} via an additivity principle.

As expected, our expression~(\ref{e:elamb}) for $e_0(\lambda)$ obeys left-right symmetry (i.e., it is invariant under the transformation $\alpha \leftrightarrow \delta$, $\beta \leftrightarrow \gamma$, $p \leftrightarrow q$ and $\lambda \leftrightarrow -\lambda$).  
Furthermore, it also demonstrates the Gallavotti-Cohen symmetry property.  For a proof of the application of this fluctuation theorem~\cite{Gallavotti95,Gallavotti95b} to stochastic particle systems see~\cite{Lebowitz99,Derrida04b}.  In the present context it arises from a particular symmetry of the Hamiltonian $\tilde{H_0}$ which we now outline.

First, note from~(\ref{e:css}) that the detailed balance condition for equilibrium (no steady-state current) is
\begin{equation}
\frac{\alpha \beta}{\gamma\delta} \left( \frac{p}{q} \right)^{L-1}=1. \label{e:detbal}
\end{equation}
This corresponds to an equilibrium steady state $|P^*_{\text{eq}}\rangle$ with fugacity $z_{\text{eq},l}=(\delta/\beta)(p/q)^{l-L}$.
In the non-equilibrium situation~(\ref{e:detbal}) is no longer satisfied but we can always write $\alpha=\alpha' \rme^{E}$ and $\gamma=\gamma' \rme^{-E}$ where detailed balance is obeyed for $\alpha'$, $\gamma'$, $\beta$, $\delta$, $p$, $q$ and $E$ is given by
\begin{equation}
\rme^{2E}=\frac{\alpha \beta}{\gamma\delta} \left( \frac{p}{q} \right)^{L-1}. \label{e:E}
\end{equation}
Physically, one thinks of $E$ as an effective external field imposed on the equilibrium model~\cite{Derrida04b}.

With these definitions we can show that $\tilde{H}$ of equation~(\ref{e:H0}) has the following symmetry property,
\begin{equation}
\tilde{H}_0^T(2E-\lambda)=(P^*_{\text{eq}})^{-1} \tilde{H}_0(\lambda) P^*_{\text{eq}} \label{e:GCFTH}
\end{equation}
where $P^*_{\text{eq}}$ is the diagonal matrix with equilibrium probabilities on the diagonal. Equation~(\ref{e:GCFTH}) means that the eigenvalues of $\tilde{H}_0(\lambda)$ and $\tilde{H}_0(2E-\lambda)$ are identical and hence, if the ground state of $\tilde{H}_0(\lambda)$ is normalizable, imposes the Gallavotti-Cohen relation~\cite{Lebowitz99} 
\begin{equation}
e_0(\lambda)=e_0(2E-\lambda)
\end{equation}
which can be straightforwardly verified on the expression~(\ref{e:elamb}).  The oft-quoted form of the fluctuation theorem, 
\begin{equation}
\frac{p_l(-j,t)}{p_l(j,t)} \sim \rme^{-2Ejt}, \label{e:GCFTj}
\end{equation}
then follows via equations~(\ref{e:pj}) and~(\ref{e:lang}).  Physically,~(\ref{e:GCFTj}) relates the probability of observing a backward current with the probability of observing a current of the same magnitude in the forward direction.  Unfortunately, such a relationship is difficult to check by simulation (or experiment) because of the exponentially small probabilities.  Nevertheless, there has been some success in this regard for other models~\cite{Evans93,Ciliberto04,Schuler05}.  As an aside, we note that the Gallavotti-Cohen fluctuation theorem reduces at equilibrium to the Green-Kubo formula and Onsager reciprocity relations---see~\cite{Gallavotti96,Lebowitz99} for details.
 
\subsection{Validity and spatial dependence}
\label{ss:cfdiss}

The expressions~(\ref{e:fug})--(\ref{e:lfug}) are general results independent of the hopping rates $w_n$.  However, the form of $w_n$ enters in the normalization of the ground state and hence the regime of validity of expression~(\ref{e:elamb}) for $e_0(\lambda)$.  The normalizing factor for $\langle 0 | 0 \rangle$ is given by the generalization of~(\ref{e:gc}) 
\begin{equation}
\sum_{n=0}^{\infty} (\tilde{z}_l z_l)^n \prod_{i=1}^{n} w_i^{-1} \label{e:gcgen}
\end{equation}
with fugacities given by (\ref{e:fug}) and (\ref{e:lfug}).  We note that, even for parameters for which the stationary state exists and there is no condensation, there may be only a restricted range of $\lambda$ for which~(\ref{e:gcgen}) is convergent and $\langle s | 0 \rangle$ is also finite. 
The characterization and physical description of current fluctuations outside this regime is a central goal of this paper.  

A related issue is the important question of whether the current fluctuations are the same across all bonds of the system, i.e., whether $e_l(\lambda)$ is independent of $l$.  Let us consider measuring the current between sites $l$ and $l+1$ in the bulk of the system.  Then
\begin{equation}
\langle \rme^{-\lambda J_l(t)} \rangle = \langle s | \rme^{-\tilde{H}_l t} |P\rangle
\end{equation}
with
\begin{eqnarray}
\fl \tilde{H}_l=- \biggl\{ \sum_{k=1, k \neq l}^{L-1} \left[  p(a_k^- a^+_{k+1} -d_k) + q (a_k^+ a_{k+1}^- - d_{k+1})\right] \nonumber \\
+ \left[  p(\rme^{-\lambda} a_l^- a^+_{l+1} -d_l) + q (\rme^{\lambda} a_l^+ a_{l+1}^- - d_{l+1})\right] \nonumber \\
+ \alpha(a_1^+ -1) + \gamma(a_1^- - d_1)  +\delta (a_L^+-1)+\beta(a^-_L-d_L) \biggr\}
\end{eqnarray}
Now, we define the operator
\begin{equation}
Y_l=\rme^{\lambda \sum_{k=1}^l d_k},
\end{equation}
and note the generic integrated commutation relation
\begin{equation}
\rme^{-\lambda(k) d_k} a_k^\pm \rme^{\lambda(k) d_k} = \rme^{\mp \lambda(k)} a_k^\pm \label{e:gencom}
\end{equation}
for $\lambda(k)$ an arbitrary function of $k$.  Using this commutation property we can show
\begin{equation}
\langle \rme^{-\lambda J_l(t)} \rangle = \langle s | Y_l \rme^{-\tilde{H}_0 t} {Y_l}^{-1} |P\rangle
\end{equation}
and hence we find, with the lowest eigenvalue $e_0(\lambda)$ of $\tilde{H}_0$, the long-time limit
\begin{equation}
\lim_{t \rightarrow \infty} \langle \rme^{-\lambda J_l(t)} \rangle = \langle s | Y_l |0 \rangle \langle 0 | {Y_l}^{-1} |P\rangle \rme^{-e_0(\lambda) t}.
\end{equation}
Here, as before, $|0\rangle$ 
is the ground-state eigenvector of $\tilde{H}_0$ and if the prefactor is finite (which now requires the additional condition that $\langle s | Y_l |0 \rangle $ is finite) then the current fluctuations across all bonds are seen to be the same (to leading order) as those going into the system, i.e., $e_l(\lambda)=e_0(\lambda)$ for all $l$.   We have verified this by explicit calculation of $e_1(\lambda)$ and $e_L(\lambda)$.  

Thus we conclude that our result~(\ref{e:elamb}) is guaranteed to describe current fluctuations across any bond $l$ but only for the range of $\lambda$ for which $\langle s|0\rangle$, $\langle 0|0\rangle$ and $\langle s | Y_l |0 \rangle $ are all finite.  
From the form of the fugacities (\ref{e:fug}),~(\ref{e:lfug})  and the normalizing factors, we see that the appropriate range of $\lambda$ will depend both on the rate parameters (i.e., $\alpha$, $\beta$, $\gamma$, $\delta$, $p$, $q$) and on the details of $w_n$.  
In order to further understand the significance of this regime of validity, and the behaviour of current fluctuations outside it, we examine in the next section the specific example of the totally asymmetric zero range process with $w_n=1$.

\section{Totally asymmetric zero-range process}
\label{s:TAZRP}

\subsection{Model and motivation}
\label{ss:TAZRPintro}

We here consider the totally asymmetric version of the model defined in section~\ref{s:ZRP}, viz.\  $\gamma=\delta=q=0$ with $p=1$ for normalization of time.  The choice $w_n=1+b/n$ is a generic model which, for periodic boundary conditions, gives condensation at high particle density for $b>2$~\cite{OLoan98,Evans00,Kafri02c,Jeon00b,Grosskinsky03,Godreche03}.  For open boundary conditions condensation can occur even for $b<2$~\cite{Levine04c}; here we study the case $b=0$ for simplicity.  

For this model, from equation~(\ref{e:zss2}), the stationary fugacities are given by
\begin{eqnarray}
z_l &= \alpha \quad \text{for} \quad l \ne L \\ 
z_L &= \frac{\alpha}{\beta}.
\end{eqnarray}
For a normalizable steady state one must have $\alpha < 1$ and $\beta > \alpha$; other cases result in the occurrence of a boundary condensate~\cite{Levine04c}.  Our aim is to study current fluctuations \emph{for all $\lambda$, $j$} in the case where the steady state is well-defined.  For definiteness we mainly consider $\alpha < \beta < 1$.

We first study analytically the toy case of a single site (section~\ref{ss:singsite}) before presenting heuristic arguments for the behaviour of current fluctuations in larger systems (section~\ref{ss:cfTAZRP}).  These arguments will be later confirmed by the Bethe ansatz approach of section~\ref{s:TASEP}.

\subsection{Current fluctuations in single-site case}                                                                                
\label{ss:singsite}

\begin{figure}
\begin{center}
\psfrag{1}[][]{1}
\psfrag{a}[][]{$\alpha$}
\psfrag{bw}[][]{$\beta$}
\includegraphics*[width=0.16\textwidth]{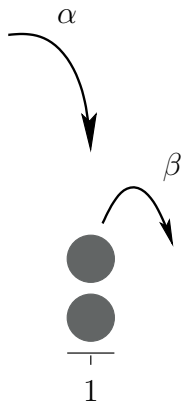}
\caption{Single-site TAZRP with $w_n=1$}
\label{f:singsite}
\end{center}
\end{figure}

It is illuminating to study the single-site (two-bond) TAZRP for which explicit calculation of $e(\lambda)$ is possible for all $\lambda$.  The model is illustrated in figure~\ref{f:singsite}; we choose $\beta > \alpha$ to ensure a well-defined steady state.

Our main results are that the current fluctuations \emph{into} the system obey,
\begin{equation}
p_0(j,t) \sim
\rme^{-t[\alpha - j +j\ln({j}/{\alpha})]}  
\end{equation}
for all currents, while those \emph{out} change in character at $j=\beta$, viz.,
\begin{equation}
p_1(j,t) \sim
\cases{
\rme^{-t[\alpha - j +j\ln({j}/{\alpha})]} & $j < \beta$ \\
\rme^{-t[\alpha - j +j\ln({j}/{\alpha})]} \times \rme^{-t[\beta - j +j\ln({j}/{\beta})]} & $j \geq \beta$. 
}
\end{equation}
The remainder of the subsection is devoted to the derivation of these expressions.  

Let us first recap the results of section~\ref{s:cf} as applied to this case.  Here $\tilde{H}_0$ of equation~(\ref{e:H0}) is given by
\begin{equation}
\tilde{H}_0=- \left[ \alpha(a_1^+ \rme^{-\lambda}-1) + \beta(a^-_1-d_1) \right] \label{e:H0TAZRPsing}
\end{equation}
which has ground-state eigenvalue  
\begin{equation}
e_0(\lambda)=\alpha(1-\rme^{-\lambda}) \label{e:e0TAZRP1}
\end{equation}
leading to a current distribution with 
\begin{equation}
\hat{e}_0(j)=\alpha - j +j\ln\frac{j}{\alpha}.
\end{equation} 
This corresponds, as should be expected, to a Poissonian distribution of particle jumps from the reservoir onto the site. According to the argument of section~\ref{ss:cfin} this result is only guaranteed to be valid if $\langle s|0\rangle$ and $\langle 0|0\rangle$ are both finite.  However, since the current entering the system is independent of the site occupation, the Poissonian distribution must hold for all $j$ and hence in this case $e_0(\lambda)$ is given by~(\ref{e:e0TAZRP1}) for \emph{all} $\lambda$.  

As we saw in section~\ref{ss:cfdiss}, the regime in which $e_1(\lambda)$ must be equal to $e_0(\lambda)$ (in other words, current fluctuations across the outgoing bond have to be given by the same Poissonian distribution) is determined by the range of $\lambda$ for which $\langle s|0\rangle$, $\langle 0|0\rangle$ and $\langle s | Y_l |0 \rangle $ are all finite.  Now, for the totally asymmetric case we find from equation~(\ref{e:lfug}) that $\tilde{z}_l=1$ so that $\langle 0 | = \langle s |$ and the first two conditions are equivalent.  Furthermore it turns out (since $z_l$ is here proportional to $\rme^{-\lambda}$) that this condition also guarantees that $\langle s | Y_l |0 \rangle $ is finite when the steady state exists.  Hence, the range of validity of $e_1(\lambda)=\alpha(1-\rme^{-\lambda})$ is simply determined by the range of normalizablity of $| 0 \rangle$.

In this single-site case, we find that $|0\rangle$ has fugacity $z_1=\alpha \rme^{-\lambda}/\beta$ so that, for $w_n=1$, the condition for normalizability is $\alpha \rme^{-\lambda} < \beta$.  Via the Legendre transformation of~(\ref{e:lang}) this gives a current condition $j<\beta$.
In order to determine the behaviour of current fluctuations out of the system when this condition is violated (i.e., $\alpha \rme^{-\lambda} \geq \beta$, $j \geq \beta$) we now explicitly calculate $e_1(\lambda)$.  In other words, we look for the lowest eigenvalue of the matrix
\begin{equation}
\tilde{H}_1 = -[\alpha(a_1^+ -1) + \beta (a_1^- \rme^{-\lambda} - d_1)] 
\end{equation}
for which the eigenvector is well-defined.\footnote{
The results of this subsection may also be obtained rigorously by calculating $\langle s | \rme^{-\tilde{H}_1 t} | P \rangle$ in integral form and examining the long-time behaviour.
}

Using the representations of creation and annihilation operators~(\ref{e:aa+}) together with the diagonal matrix $d_1$ we see that the right eigenvector corresponding to eigenvalue $A$ satisfies
\begin{equation}
\left(
\begin{array}{ccccc}
\alpha & -\beta \rme^{-\lambda} & 0 & 0 & \ldots \\
-\alpha & \alpha+\beta & -\beta \rme^{-\lambda} & 0 & \ldots \\
0 & -\alpha & \alpha+\beta & -\beta \rme^{-\lambda} & \ldots \\
0 & 0 & -\alpha & \alpha+\beta & \ldots \\
\ldots & \ldots & \ldots & \ldots & \ldots
\end{array}
\right)
\left(
\begin{array}{c}
b_0 \\
b_1 \\
b_2 \\
b_3 \\
\ldots 
\end{array}
\right)
= A
\left(
\begin{array}{c}
b_0 \\
b_1 \\
b_2 \\
b_3 \\
\ldots 
\end{array}
\right)
\end{equation}
We thus obtain a mapping for the elements of the eigenvector
\begin{equation}
b_{n+1}=\frac{-\alpha b_{n-1} + (\alpha + \beta - A) b_n}{\beta \rme^{-\lambda}}
\end{equation}
with initial condition
\begin{equation}
b_1 = \left( \frac{\alpha - A}{\beta \rme^{-\lambda}} \right) b_0.
\end{equation}
This mapping can be written as a transfer matrix equation
\begin{equation}
\left(
\begin{array}{c}
b_{n+1} \\
b_n
\end{array}
\right)
=
\left(
\begin{array}{cc}
\frac{\alpha+\beta-A}{\beta \rme^{-\lambda}} & \frac{-\alpha}{\beta \rme^{-\lambda}} \\
1 & 0 
\end{array}
\right)
\left(
\begin{array}{c}
b_n \\
b_{n-1}
\end{array}
\right)
\end{equation}
where the transfer matrix has eigenvalues
\begin{equation}
\mu_\pm = \frac{\alpha + \beta - A \pm \sqrt{(\alpha + \beta - A)^2 - 4 \alpha \beta \rme^{-\lambda}}}{2 \beta \rme^{-\lambda}}.
\end{equation}

Similarly the elements of the corresponding left eigenvector $(a_0,a_1,a_2,a_3,\ldots)$ of $\tilde{H}_1$ are given by the mapping
\begin{eqnarray}
a_1 &= \left( \frac{\alpha - A}{\alpha} \right) a_0 \\
a_{n+1}&=\frac{-\beta \rme^{-\lambda} a_{n-1} + (\alpha + \beta - A) a_n}{\alpha}
\end{eqnarray}
corresponding to a transfer matrix with eigenvalues
\begin{equation}
\nu_\pm = \frac{\alpha + \beta - A \pm \sqrt{(\alpha + \beta - A)^2 - 4 \alpha \beta \rme^{-\lambda}}}{2 \alpha}.
\end{equation}

From the properties of these transfer matrix eigenvalues we can determine the large $n$ behaviour of $a_n$ (and $b_n$).  Specifically, the limit of the ratio $ a_n/a_{n-1}$ is given by the largest magnitude eigenvalue \emph{except} in the case that the initial ratio $ a_1 / a_0$ coincides with the smaller eigenvalue.  Now, if $\lim_{n\to\infty}|(a_n b_n)/(a_{n-1} b_{n-1})| > 1$, then the series $a_0 b_0 + a_1 b_1 + a_2 b_2 + \dots a_n b_n \dots$ is divergent (D'Alembert ratio test) and hence the eigenvector of $\tilde{H}_1$ with eigenvalue $A$ cannot be normalized.  Such an analysis leads to the conclusion that the lowest eigenvalue for which the corresponding eigenvector can be defined is given by\footnote{One can readily check that in both cases $\langle s | 1 \rangle$ is finite with $| 1 \rangle$ the right eigenvector corresponding to $e_1(\lambda)$.}
\begin{equation}
e_1(\lambda) =
\cases{
\alpha (1 - \rme^{-\lambda}) & $\alpha \rme^{-\lambda} < \beta$ \\
\alpha + \beta - 2 \sqrt{\alpha \beta \rme^{-\lambda}}  & $\alpha \rme^{-\lambda} \geq \beta$.
}
\end{equation}
Using the Legendre transformation~(\ref{e:lang}) this gives
\begin{equation}
\hat{e}_1(j,t) = 
\cases{
\alpha - j +j\ln\frac{j}{\alpha} & $j < \beta$ \\
\alpha - j +j\ln\frac{j}{\alpha} + \beta - j +j\ln\frac{j}{\beta}  & $j \geq \beta$.
}
\end{equation}
In other words, for currents greater than $\beta$ the current fluctuations out of the site are different to those into the site.  This behaviour can be explained heuristically as follows.

For $j < \beta$, the current into the site is smaller than the average attempt rate out.  This means that hopping across the two bonds is not independent and there is no chance for particles to build-up on the site---all particles which arrive there move almost immediately out of the system.  Hence, for $j<\beta$, the limiting factor in observing a current fluctuation with $j_1=j$ is just the input current $j_0$, i.e, $\mathrm{Prob}(j_1=j,t) \sim \mathrm{Prob}(j_0 \geq j,t)$ and we find that $\hat{e}_1(j,t)=\hat{e}_0(j,t)=\alpha - j +j\ln({j}/{\alpha})$.  

In contrast, for $j > \beta$, the current into the site is greater than the attempt rate out leading to a temporary build-up of particles which we dub an ``instantaneous condensate''.  (We emphasize that for the parameters we study, $\alpha < \beta$, there is no permanent condensation, the temporary build-up of particles occurs only in those unlikely realizations where $j>\beta$.)  In this case, at large times the probability of finding at least one particle on the site is essentially unity, so the two bonds behave as independent Poissonian processes (with means $\alpha$ and $\beta$ respectively).  Hence, for $j>\beta$, $\mathrm{Prob}(j_1=j,t)$ is given by the product of $\mathrm{Prob}(j_0\geq j,t)$ with the independent Poissonian probability for observing a current $j$ across the outgoing bond.  This gives $\hat{e}_1(j,t)=\alpha - j +j\ln({j}/{\alpha})+\beta - j +j\ln( {j}/{\beta}) $.

\subsection{Current fluctuations in $L$-site case}
\label{ss:cfTAZRP}

Based on this picture of ``instantaneous condensates'' we propose that current fluctuations across the different bonds in an $L$-site TAZRP, with $w_n=1$ and $\alpha < \beta < 1$, are given by the following expressions:
\begin{itemize}
\item Input bond
\begin{equation}
p_0(j,t) \sim
\rme^{-t[\alpha - j +j\ln({j}/{\alpha})]}.  \label{e:q0}
\end{equation}

\item Bulk bonds, $l\neq 0,L$
\begin{equation}
p_l(j,t) \sim
\cases{
\rme^{-t[\alpha - j +j\ln({j}/{\alpha})]} & $j < 1$ \\
\rme^{-t[\alpha - j +j\ln({j}/{\alpha})]} \times \rme^{-t(1 - j +j\ln j)l} & $j \geq 1$.   \label{e:ql}
}
\end{equation}
Note the $l$-dependence here!

\item Output bond
 \begin{equation}
\fl p_L(j,t) \sim
\cases{
\rme^{-t[\alpha - j +j\ln({j}/{\alpha})]} & $j < \beta$ \\
\rme^{-t[\alpha - j +j\ln({j}/{\alpha})]} \times \rme^{-t[(\beta - j +j\ln ({j}/{\beta})]} & $\beta \leq j < 1$ \\
\rme^{-t[\alpha - j +j\ln({j}/{\alpha})]} \times \rme^{-t(1 - j +j\ln j)(L-1)} \times \rme^{-t[\beta - j +j\ln({j}/{\beta})]} & $j \geq 1$. \label{e:qL} 
}
\end{equation}

\end{itemize}
In the next section we argue that these expressions can be rigorously obtained via a Bethe ansatz approach.  First we present here some general arguments for their form and significance.

The low current behaviour can be directly obtained from the analysis of section~\ref{s:cf}.  The lowest eigenvalue of 
\begin{equation}
\tilde{H}_0=- \left[ \sum_{l=1}^{L-1} p(a_l^- a^+_{l+1} -d_l) + \alpha(a_1^+ \rme^{-\lambda}-1) + \beta(a^-_L-d_L) \right]. \label{e:H0TAZRP}
\end{equation}
is 
\begin{equation}
e_0(\lambda)=\alpha(1-\rme^{-\lambda}) \label{e:e0TAZRP}
\end{equation}
for any $L$, again giving a Poissonian distribution of current into the system, i.e.,
\begin{equation}
\hat{e}_0(j)=\alpha - j +j\ln\frac{j}{\alpha}.
\end{equation} 

\begin{figure}
\begin{center}
\psfrag{0}[Bl][Bl]{0}
\psfrag{a}[Bl][Bl]{$\alpha$}
\psfrag{b1}[Bc][Bc]{$\beta$}
\psfrag{b}[Cl][Cl]{$\beta$}
\psfrag{e}[Bl][Bl]{$\rme^\lambda$}
\psfrag{j}[Bl][Bl]{$j$}
\psfrag{1}[Cl][Cl]{1}
\psfrag{I}[][]{I}
\psfrag{II}[][]{II}
\psfrag{III}[][]{III}
\mbox{\subfigure[]{\includegraphics*[width=0.3\textwidth]{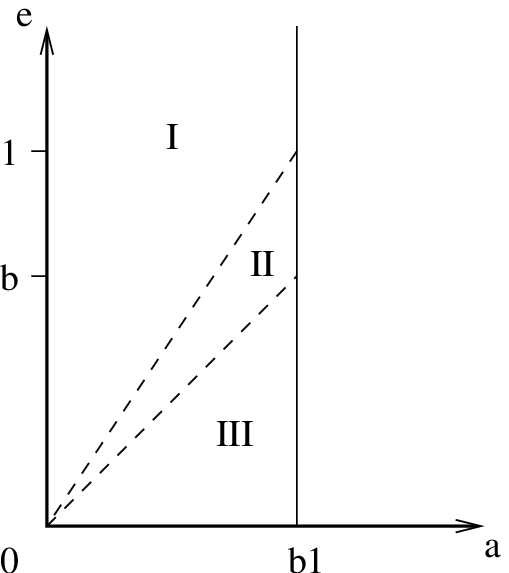}}\qquad
\subfigure[]{\includegraphics*[width=0.3\textwidth]{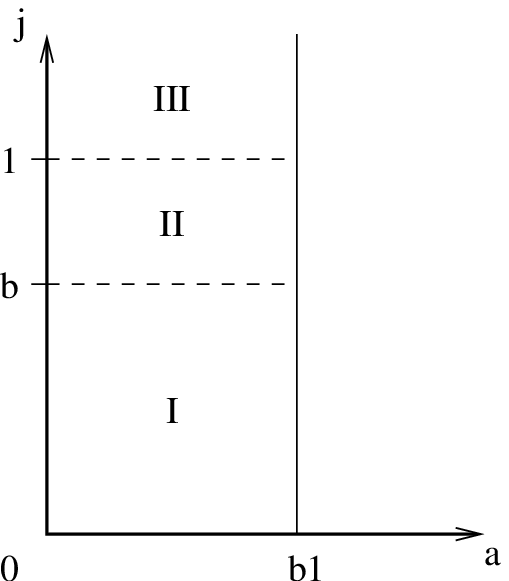}}}
\caption{Phase diagram of TAZRP (with $w_n=1$) for case $\beta < 1$ in (a) $\alpha$\,--\,$\rme^\lambda$ plane and (b) $\alpha$\,--\,$j$ plane.  Phases I, II and III correspond to the different regimes of current fluctuations described in the text.  No steady state exists to the right of the solid line.}
\label{f:pd}
\end{center}
\end{figure}

The ground state $|0\rangle$ of $\tilde{H}_0$ corresponding to eigenvalue $e_0(\lambda)$ diverges on site $L$ for $\alpha \rme^{-\lambda} \geq \beta$ ($j \geq \beta$) and on all sites for $\alpha \rme^{-\lambda} \geq 1$ ($j \geq 1$).  For current fluctuations above these limits we argue that the divergence of this ground state indicates the temporary build-up of particles---``instantaneous condensates''---in this sense it is a precursor of the condensation which would occur for $\beta < \alpha$ or $\alpha > 1$. This causes the spatial homogeneity of the current fluctuations to be broken.  In summary, we expect to find 3 different regimes:
\begin{enumerate}[I.]
\item {\boldmath$\lambda > \ln(\alpha/\beta)$, $j < \beta$\unboldmath:  Current fluctuations across all bonds are identical with $e_l(\lambda)=e_0(\lambda)=\alpha(1-\rme^{-\lambda})$ and $\hat{e}_l(j)=\hat{e}_0(j)=\alpha - j +j\ln({j}/{\alpha})$ for all $l$.}

\item {\boldmath$\ln(\alpha/\beta) \geq \lambda > \ln\alpha$, $\beta \leq j < 1$\unboldmath:  In this case the eigenvector corresponding to $e_0(\lambda)$ is not normalizable for site $L$.  Physically, we argue that for $j>\beta$ the current into site $L$ is larger than the attempt rate out leading to a temporary build-up of particles on site $L$.
The probability of observing a current fluctuation with $j_L=j$ is then given by the probability of observing $j_{L-1} \geq j$ multiplied by the independent Poissonian probability (mean $\beta$) of particles at site $L$ exiting the system with rate $j$.  For all other sites the ground state is well-defined, behaving like a system of size $L-1$ with effective exit parameter $\beta'=1$.  Hence, for $l \neq L$, $\hat{e}_l(j)=\hat{e}_0(j)$.}

\item {\boldmath$\lambda \leq \ln\alpha$, $j\geq  1$\unboldmath: Here the eigenvector corresponding to $e_0(\lambda)$ is ill-defined on all sites and hence we expect ``instantaneous condensates'' on all sites.  Each bond then acts as an independent Poisson process.  A particular feature of this totally asymmetric model is that the current through a given bond is dependent only on the bonds to the left.  Hence, for $j>1$, $\mathrm{Prob}(j_l=j,t)$ is a product of $l+1$ Poissonian probabilities and the distribution of current fluctuations through every bond is different.

}

\end{enumerate}
For fixed $\beta<1$, this behaviour can be summarized in the 2D phase diagrams of figure~\ref{f:pd}.

For the alternative case $\alpha < 1 < \beta$ one expects instantaneous condensates on sites 1 to $L-1$ for $j>1$ and on all sites for $j>\beta$.  The probability distribution for current fluctuations out of the system will be accordingly altered to
\begin{equation}
\fl p_L(j,t) \sim
\cases{
\rme^{-t[\alpha - j +j\ln({j}/{\alpha})]} & $j \leq 1$ \\
\rme^{-t[\alpha - j +j\ln({j}/{\alpha})]} \times \rme^{-t(1 - j +j\ln j)(L-1)} & $1 \leq j \leq \beta$ \\
\rme^{-t[\alpha - j +j\ln({j}/{\alpha})]} \times \rme^{-t(1 - j +j\ln j)(L-1)} \times \rme^{-t[\beta - j +j\ln({j}/{\beta})]} & $j \geq \beta$. \label{e:qLb} 
}
\end{equation}

\section{Application of Bethe ansatz results}
\label{s:TASEP}

\subsection{Mapping to exclusion process}
\label{ss:prelim}

In this section we demonstrate how the application of new Bethe ansatz results~\cite{Rakos05} for the totally asymmetric exclusion process with particle-dependent hopping rates provides an alternative analytical approach to rigorously derive the TAZRP results of the previous section.  In particular, we wish to recover equations~(\ref{e:q0})--(\ref{e:qL}) for the site-dependent probability of observing current fluctuations with a given $j$.

\begin{figure}
\begin{center}
\psfrag{0}[][]{1}
\psfrag{1}[][]{2}
\psfrag{2}[][]{3}
\psfrag{3}[][]{4}
\psfrag{4}[][]{5}
\psfrag{5}[][]{6}
\psfrag{L-2}[][]{$L\!-\!1$}
\psfrag{L-1}[][]{$L$}
\psfrag{L}[][]{$L+1$}
\psfrag{a}[][]{$\alpha$}
\psfrag{d}[][]{$\delta$}
\psfrag{b}[][]{$\beta$}
\psfrag{c}[][]{$\gamma$}
\psfrag{p}[][]{$p$}
\psfrag{q}[][]{$q$}
\includegraphics*[width=1.0\textwidth]{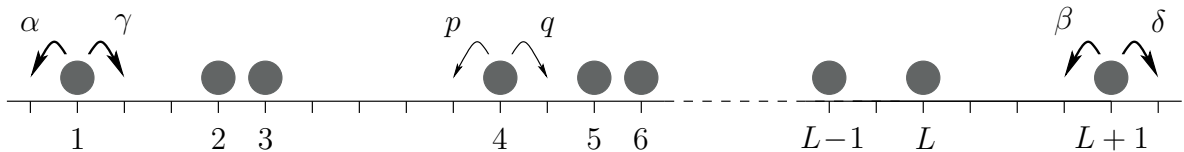}
\caption{Exclusion model corresponding to zero-range process of figure~\ref{f:ZRP} with $w_n=1$.  Bulk particles hop to~\emph{vacant} nearest-neighbour sites on the left (right) with rate $p$ ($q$).  Particles $1$ and $L+1$ have different hopping rates corresponding to the boundary processes of the ZRP.}
\label{f:excl}
\end{center}
\end{figure}

The general ZRP can be mapped to an exclusion model (in which lattice sites cannot be multiply occupied) by representing each bond as a particle and the occupation of the site between a pair of bonds as the number of interparticle vacancies.  For example, the particle configuration shown in the $L$-site ZRP of figure~\ref{f:ZRP} (with each bond labelled by its lefthand site) is equivalent to the exclusion picture of figure~\ref{f:excl} with $L+1$ particles---note that with our numbering convention the $l$th bond of the ZRP maps to the $(l+1)$th particle in the exclusion process.
The hopping of a ZRP particle in a rightward (leftward) direction across a given bond, then corresponds to the equivalent exclusion particle moving one lattice site \emph{left} (\emph{right}).  For the case $w_n=1$ (the pure chipping process~\cite{Majumdar98}) the hopping attempt rates are independent of the particle separation leading to the (asymmetric) simple exclusion process with parameters assigned as in figure~\ref{f:excl}.

We emphasize that the ZRP on a finite open lattice maps to an exclusion process with a fixed number of particles on an \emph{infinite} lattice where the endmost particles have different hopping rates from the others.  An initially empty lattice in the ZRP corresponds to the particles in the exclusion process all occupying adjacent sites (no interparticle vacancies).

In particular, the open-boundary totally asymmetric zero range process (TAZRP) of section~\ref{s:TAZRP} maps to the totally asymmetric exclusion process (TASEP) on an infinite lattice.  The pure TASEP (particle-independent hopping rates) has been extensively studied, for both periodic and open boundary conditions, and the steady state solved by a variety of means including recursive techniques~\cite{Schutz93c}, a steady-state operator algebra formulation~\cite{Derrida93b} and application of the Bethe ansatz to the corresponding quantum-spin system~\cite{Gwa92b}.  Recently R{\'a}kos and Sch{\"u}tz~\cite{Rakos05} have extended the Bethe approach for calculating current fluctuations to the general case of particle-dependent hopping rates.  We now summarize the result of that calculation, reordering the numbering of particles in order to make the connection with the TAZRP more transparent.

Consider an infinite one-dimensional lattice containing $N$ particles with \emph{left} hopping rates $v_1, v_2, \dots, v_N$ (from left to right).  If the particles are initially located at sites $y_1, y_2, \dots, y_N$ then the probability $P(\bi{x},\bi{y},t)$ of finding them at time $t$ on sites $x_1, x_2, \dots, x_N$ (with lattice coordinate increasing to the left and $x_l \geq y_l\ \forall l$) can be written in the determinant form
\begin{eqnarray}
P(\bi{x},\bi{y},t)=\prod_{i=1}^N(\rme^{-tv_i}{v_i}^{x_i-y_i}) \times \\
\left|
\begin{array}{cccc}
F_{1,1}(x_1-y_1,t) & F_{1,2}(x_2-y_1,t) & \dots & F_{1,N}(x_N-y_1,t) \\
F_{2,1}(x_1-y_2,t) & F_{2,2}(x_2-y_2,t) & \dots & F_{2,N}(x_N-y_2,t) \\
\dots & \dots & \dots & \dots \\
F_{N,1}(x_1-y_N,t) & F_{N,2}(x_2-y_N,t) & \dots & F_{N,N}(x_N-y_N,t)
\end{array}
\right| \label{e:BAresult}
\end{eqnarray}
with $F$-functions defined as
\begin{equation}
F_{r,s}(x,t)=\frac{1}{2\pi \rmi} \oint \rme^{t/z} z^{x-1} \prod_{i=r+1}^{N} (1-v_i z) \prod_{i=s+1}^{N} (1-v_i z)^{-1} \, \rmd z \label{e:Fdef}
\end{equation}
where the integral is to be taken along a circle of radius $\epsilon$ around the origin of the complex plane.  These functions obey the following identities \begin{eqnarray}
\sum_{x=x_1}^{x_2} {v_s}^{x} F_{r,s}(x,t) &= {v_s}^{x_1} F_{r,s-1}(x_1,t)-{v_s}^{x_2+1}F_{r,s-1}(x_2+1,t) \\
\sum_{x=x_1}^{x_2} {v_{r+1}}^{x} F_{r,s}(x,t) &= {v_{r+1}}^{x_1} F_{r+1,s}(x_1,t)-{v_{r+1}}^{x_2+1}F_{r+1,s}(x_2+1,t)
\end{eqnarray}
with important special cases for $x_1=x_2=x$
\begin{eqnarray}
F_{r,s}(x,t) &= F_{r,s-1}(x,t)-v_s F_{r,s-1}(x+1,t) \\
F_{r,s}(x,t) &= F_{r+1,s}(x,t)-v_{r+1} F_{r+1,s}(x+1,t).
\end{eqnarray}

In the next subsection we show how the result (\ref{e:BAresult}) can be applied to obtain the current fluctuations in the TAZRP.  Note that for brevity we will write $F(x)$ as a shorthand for $F(x,t)$.

\subsection{Determinant solution for current fluctuations in TAZRP}
\label{ss:caseTAZRP}


As outlined above, we map the $L$ site TAZRP to an $(L+1)$-particle TASEP where the particles hop with rates $v_1=\alpha$, $v_{L+1}=\beta$ and $v_{l\neq 1,L+1}=p=1$.  To study current fluctuations across the $l$th bond of the TAZRP we need to know about the number of jumps made by the corresponding exclusion particle.  Importantly, for this totally asymmetric model, this is independent of all the particles behind and thus we need only consider the first $l+1$ particles.  For simplicity we will start from the initial configuration of adjacent exclusion particles (equivalent to an initially empty lattice in the ZRP); as previously discussed, we expect the long-time decay of the current fluctuations to be independent of the specific initial particle configuration. 

Now we can use the Bethe ansatz solution to obtain the distribution of the distance travelled by the $(l+1)$th particle
\begin{eqnarray}
\fl \mathrm{Prob}(x_1 \geq x+l, x_2 \geq x+l-1, \dots, x_l \geq x+1, x_{l+1}=x | \nonumber \\ y_1=l, y_2=l-1,
\dots y_l=1, y_{l+1}=0,t) \nonumber \\
\fl \eql \sum_{x_1=x+l}^\infty  \sum_{x_2=x+l-1}^{x_1-1} \dots \sum_{x_l=x+1}^{x_{l-1}-1} P(\{x_1,x_2 \dots x_l, x\},\bi{y},t) \\
\fl \eql \prod_{i=1}^{l+1}(\rme^{-tv_i}{v_i}^{x}) \times  \nonumber \\
\fl \left|
\begin{array}{ccccc}
F_{1,0}(x) & F_{1,1}(x-1) & \dots & F_{1,l-1}(x-l+1) & F_{1,l+1}(x-l) \\
F_{2,0}(x+1) & F_{2,1}(x) & \dots & F_{2,l-1}(x-l+2) & F_{2,l+1}(x-l+1) \\
\dots & \dots & \dots & \dots & \dots \\
F_{l+1,0}(x+l) & F_{l+1,1}(x+l-1) & \dots & F_{l+1,l-1}(x+1) & F_{l+1,l+1}(x)
\end{array}
\right| \\
\fl \eql \prod_{i=1}^{l+1}(\rme^{-tv_i}{v_i}^{x}) \times \nonumber \\
\fl \left|
\begin{array}{ccccc}
F_{l+1,0}(x) & F_{l+1,0}(x-1) & \dots & F_{l+1,0}(x-l+1) & F_{l+1,l+1}(x-l) \\
F_{l+1,0}(x+1) & F_{l+1,0}(x) & \dots & F_{l+1,0}(x-l+2) & F_{l+1,l+1}(x-l+1) \\
\dots & \dots & \dots & \dots & \dots \\
F_{l+1,0}(x+l) & F_{l+1,0}(x+l-1) & \dots & F_{l+1,0}(x+1) & F_{l+1,l+1}(x) 
\end{array}
\right|
\end{eqnarray}
where the intermediate steps use determinant manipulation together with $F$-function identities given in the preceding subsection~\cite{Rakos04b}.

To obtain an explicit expression for the current distribution across the $l$th TAZRP bond we set $x=J=jt$ and drop the redundant subscript to yield 
\begin{eqnarray}
\fl p_l(j,t) = \prod_{i=1}^{l+1} \rme^{-t(v_i-j\ln{v_i})} \times \nonumber \\
\fl { \left| \begin{array}{ccccc}
{ D_{0}(jt,t)} & { D_{0}(jt-1,t)} & \dots & { D_{0}(jt-l+1,t)} & { D_{l+1}(jt-l,t)} \\
{ D_{0}(jt+1,t)} & { D_{0}(jt,t)} & \dots & { D_{0}(jt-l+2,t)} &  { D_{l+1}(jt-l+1,t)} \\
{ \dots} & { \dots} & { \dots} & {\dots} & { \dots} \\
{ D_{0}(jt+l,t)} & { D_{0}(jt+l-1,t)} & \dots & { D_{0}(jt+1,t)} & { D_{l+1}(jt,t)} \label{e:detsol}
\end{array} \right| }
\end{eqnarray}
with elements
\begin{equation}
D_{s}(x,t)=
\frac{1}{2\pi \rmi} \oint  \rme^{{t}/{z}} z^{x-1} \prod_{i=s+1}^{l+1} (1-v_i z)^{-1}\, \rmd z. 
\end{equation}
For the TAZRP with initially empty lattice this expression is exact for all times; we expect the long-time limit to give the initial-configuration-independent expressions suggested previously~(\ref{e:q0})--(\ref{e:qL}).

To understand how different regimes of current fluctuations emerge from this determinant solution, let us examine the case of a two-site TAZRP (equivalent to a three-particle TASEP).  As in section~\ref{s:TAZRP}, we take $\alpha<\beta<1$.  

We begin by considering the current fluctuations across the 0th bond of the TAZRP.  In this case the determinant is just a single integral
\begin{equation}
p_0(j,t) = \rme^{-t(\alpha-j\ln{\alpha})} \frac{1}{2\pi \rmi} \oint  \rme^{{t}/{z}} z^{jt-1} \, \rmd z 
\end{equation}
and a saddle-point approximation leads straightforwardly to the asymptotic behaviour
\begin{equation}
p_0(j,t) \sim
\rme^{-t(\alpha - j +j\ln\frac{j}{\alpha})}  \label{e:q20}
\end{equation}
in agreement with our previous arguments (recall that the current into the system is always described by a Poissonian distribution).

For current fluctuations across the 1st bond of the TAZRP (i.e, between sites 1 and 2) we need a $2\times 2$ determinant
\begin{equation}
\fl p_1(j,t)=\rme^{-t(\alpha-j\ln{\alpha}+1)}
\left|
\begin{array}{cc}
\frac{1}{2\pi \rmi} \oint  \rme^{{t}/{z}} z^{jt-1} (1- z)^{-1} (1- \alpha z)^{-1} \, \rmd z & \frac{1}{2\pi \rmi} \oint  \rme^{{t}/{z}} z^{jt-2}\, \rmd z \\
\frac{1}{2\pi \rmi} \oint \rme^{{t}/{z}} z^{jt} (1- z)^{-1} (1- \alpha z)^{-1} \, \rmd z  & \frac{1}{2\pi \rmi} \oint  \rme^{{t}/{z}} z^{jt-1}\, \rmd z.  \\
\end{array}
\right|
\end{equation}
Again we argue that the long-time behaviour can be extracted by saddle-point analysis of the individual elements.  However, more care is now needed because of the poles in the integrands of the $D_0$ functions.  Specifically, we note that these poles are at $z=1$ and $z=\alpha^{-1}$ while the saddle-point is located at $z=j^{-1}$.

For $j>1$, both poles lie outside the saddle-point contour.  In order to obtain the leading-order non-zero term in the determinant it is necessary to expand the $t$-independent factors in the integrands to second order about the saddle-point.  This gives the long-time behaviour
\begin{equation}
p_1(j,t) \sim \rme^{-t[\alpha - j +j\ln({j}/{\alpha})]} \times \rme^{-t(1 - j +j\ln j)}. \label{e:q21a}
\end{equation}

In contrast, for $j<1$, in deforming the contour through the saddle-point we include the pole at $z=1$.  Subtracting the residue at this pole from the saddle-point approximation for each $D_0$ integral yields the changed leading-order behaviour for the determinant
\begin{equation}
p_1(j,t) \sim \rme^{-t[\alpha - j +j\ln({j}/{\alpha})]}. \label{e:q21b}
\end{equation}
For $j<\alpha$ the pole at $z=\alpha^{-1}$ is also inside the contour.  However it turns out that this results in the addition of a sub-leading term and the long-time limit is unchanged.  Hence, for the full range of $j$, we recover the current fluctuation distribution predicted in section~\ref{s:TAZRP} [cf.\ equation~(\ref{e:ql}) with $l=1$].

To determine current fluctuations across the 2nd bond of our two-site TAZRP (i.e., the current fluctuations from site 2 out of the system), we must add a third particle with hopping rate $\beta$ to our TASEP picture and evaluate the $3\times3$ determinant given by~(\ref{e:detsol}) with $l=2$.  Here a similar saddle-point analysis with careful treatment of the poles at $z=\alpha^{-1}$, $z=1$ and $z=\beta^{-1}$ leads to
\begin{equation}
\fl p_2(j,t) \sim
\cases{
\rme^{-t[\alpha - j +j\ln({j}/{\alpha})]} & $j < \beta$ \\
\rme^{-t[\alpha - j +j\ln({j}/{\alpha})]} \times \rme^{-t[\beta - j +j\ln ({j}/{\beta})]} & $\beta \leq j < 1$ \\
\rme^{-t[\alpha - j +j\ln({j}/{\alpha})]} \times \rme^{-t(1 - j +j\ln j)} \times \rme^{-t[\beta - j +j\ln ({j}/{\beta})]} & $j \geq 1$, 
} \label{e:q22}
\end{equation}
in agreement with equation~(\ref{e:qL}) for the case $L=2$.

\begin{figure}
\begin{center}
\psfrag{j}[][]{$j$}
\psfrag{e}[][]{$\hat{e}_2(j)$}
\psfrag{ 0b}[Tc][Tc]{\footnotesize{0.0}}
\psfrag{ 0.2b}[Tc][Tc]{\footnotesize{0.2}}
\psfrag{ 0.4b}[Tc][Tc]{\footnotesize{0.4}}
\psfrag{ 0.6b}[Tc][Tc]{\footnotesize{0.6}}
\psfrag{ 0.8b}[Tc][Tc]{\footnotesize{0.8}}
\psfrag{ 1}[Tc][Tc]{\footnotesize{1.0}}
\psfrag{ 0}[Cr][Cr]{\footnotesize{0.0}}
\psfrag{ 0.1}[Cr][Cr]{\footnotesize{0.1}}
\psfrag{ 0.2}[Cr][Cr]{\footnotesize{0.2}}
\psfrag{ 0.3}[Cr][Cr]{\footnotesize{0.3}}
\psfrag{ 0.4}[Cr][Cr]{\footnotesize{0.4}}
\psfrag{ 0.5}[Cr][Cr]{\footnotesize{0.5}}
\psfrag{ 0.6}[Cr][Cr]{\footnotesize{0.6}}
\psfrag{ 0.7}[Cr][Cr]{\footnotesize{0.7}}
\psfrag{ 0.8}[Cr][Cr]{\footnotesize{0.8}}
\psfrag{ 0.9}[Cr][Cr]{\footnotesize{0.9}}
\psfrag{h(x)}[Cl][Cl]{\footnotesize{Long-time limit~(\ref{e:q22})}}
\psfrag{"threeparticles100.dat"}[Cl][Cl]{\footnotesize{Determinant solution, $t=100$}}
\psfrag{"threeparticles300.dat"}[Cl][Cl]{\footnotesize{Determinant solution, $t=300$}}
\psfrag{"threeparticles1000.dat"}[Cl][Cl]{\footnotesize{Determinant solution, $t=1000$}}
\psfrag{"LPPpapc_100_2x9_2.dat"}[Cl][Cl]{\footnotesize{Simulation, $t=100$}}
\includegraphics*[width=1.0\textwidth]{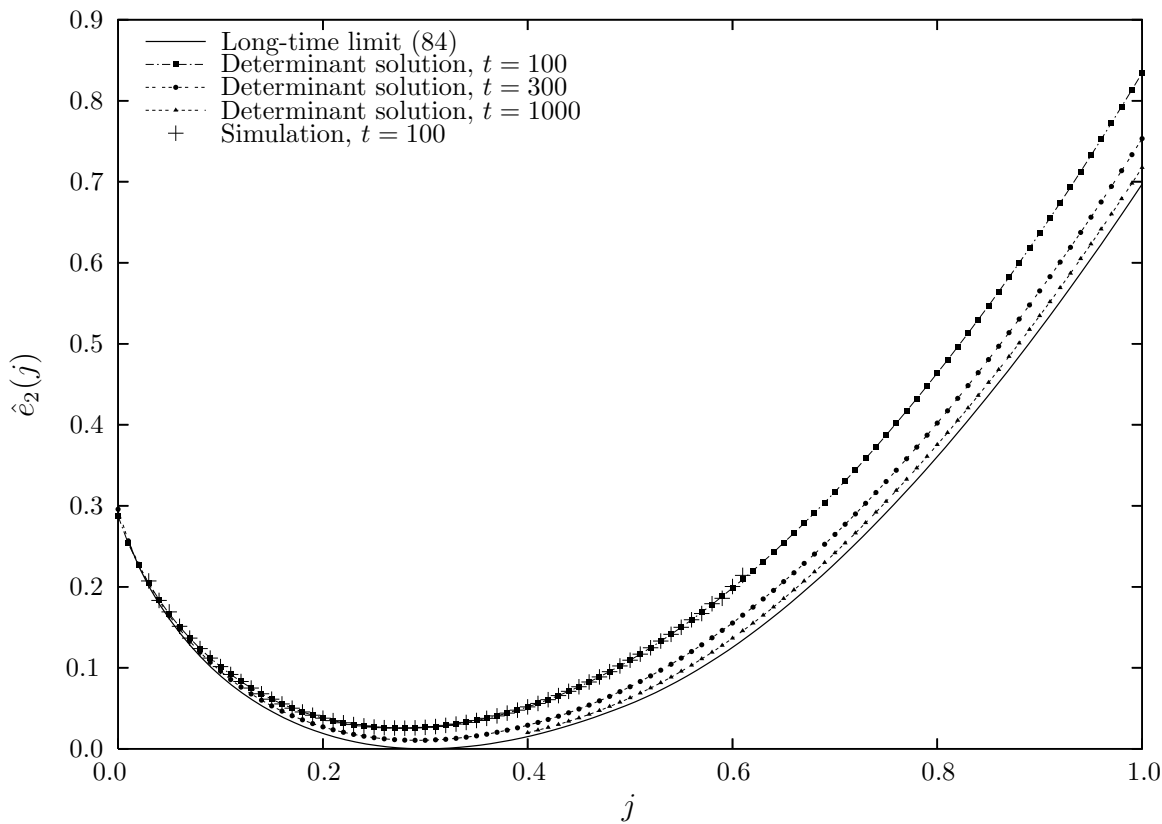}
\caption{Current fluctuations across the 2nd bond (out of the system) in a 2-site TAZRP with $\alpha=0.3$, $p=1$ and $\beta=0.5$. Graph shows $\hat{e}_2(j)=\lim_{t\rightarrow \infty} (1/t) \ln p_2(j,t)$.  Solid symbols on dashed lines result from numerical evaluation of the determinant solution at times $t=100$, 300 and 1000.  Crosses are simulation results for $t=100$.  For increasing times, distributions converge on the long-time limit of equation~(\ref{e:q22}) shown by the solid line.}
\label{f:num1}
\end{center}
\end{figure}

The validity of this saddle-point approach is confirmed by numerical evaluation of the determinants.  In figure~\ref{f:num1} 
we show the result for the second bond for several different times.  Convergence to the long-time limit of~(\ref{e:q22}) is clearly seen.  Good agreement is also shown with simulation results for current fluctuations in the totally asymmetric zero-range process with empty initial lattice.\footnote{
For efficient simulation we utilized the exact mapping of the TAZRP to a last passage percolation problem~\cite{Praehofer02}.
}
For other initial configurations we found slightly different curves for small times but, as expected, the same limiting distribution for long times.  Unfortunately, since the chance of seeing a current fluctuation away from the average becomes exponentially more unlikely with time, obtaining large deviation simulation data at long times is very computationally intensive.

\begin{figure}
\begin{center}
\psfrag{j}[][]{$j$}
\psfrag{e}[][]{$\hat{e}(j)$}
\psfrag{ 0b}[Tc][Tc]{\footnotesize{0.0}}
\psfrag{ 0.2}[Tc][Tc]{\footnotesize{0.2}}
\psfrag{ 0.4}[Tc][Tc]{\footnotesize{0.4}}
\psfrag{ 0.6}[Tc][Tc]{\footnotesize{0.6}}
\psfrag{ 0.8}[Tc][Tc]{\footnotesize{0.8}}
\psfrag{ 1b}[Tc][Tc]{\footnotesize{1.0}}
\psfrag{ 1.2}[Tc][Tc]{\footnotesize{1.2}}
\psfrag{ 1.4}[Tc][Tc]{\footnotesize{1.4}}
\psfrag{ 0}[Cr][Cr]{\footnotesize{0.0}}
\psfrag{ 0.5}[Cr][Cr]{\footnotesize{0.5}}
\psfrag{ 1}[Cr][Cr]{\footnotesize{1.0}}
\psfrag{ 1.5}[Cr][Cr]{\footnotesize{1.5}}
\psfrag{ 2}[Cr][Cr]{\footnotesize{2.0}}
\psfrag{ 2.5}[Cr][Cr]{\footnotesize{2.5}}
\psfrag{f(x)}[Cl][Cl]{\footnotesize{Long-time limit 0th bond}}
\psfrag{g(x)}[Cl][Cl]{\footnotesize{Long-time limit 1st bond}}
\psfrag{h(x)}[Cl][Cl]{\footnotesize{Long-time limit 2nd bond}}
\psfrag{"oneparticle1000.dat"}[Cl][Cl]{\footnotesize{Determinant solution 0th bond, $t=1000$}}
\psfrag{"twoparticles1000.dat"}[Cl][Cl]{\footnotesize{Determinant solution 1st bond, $t=1000$}}
\psfrag{"threeparticles1000.dat"}[Cl][Cl]{\footnotesize{Determinant solution 2nd bond, $t=1000$}}
\includegraphics*[width=1.0\textwidth]{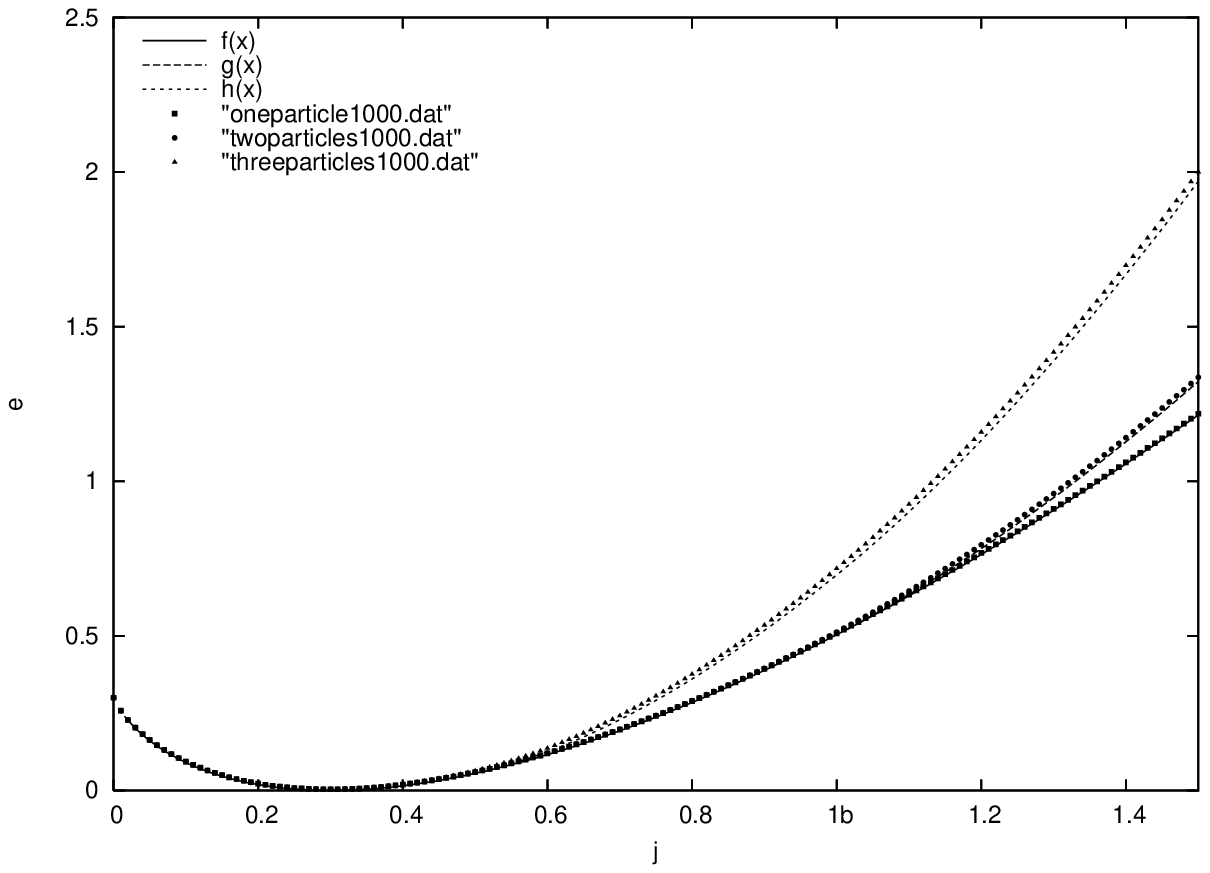}
\caption{Current fluctuations across the three different bonds of a 2-site TAZRP with $\alpha=0.3$, $p=1$ and $\beta=0.5$. Graph shows $\hat{e}_l(j)=\lim_{t\rightarrow \infty} (1/t) \ln p_l(j,t)$.  Solid symbols result from numerical evaluation of the determinant solution at time $t=1000$.  Lines are the long-time limits given by equations~(\ref{e:q20}) and~(\ref{e:q21a})--(\ref{e:q22}).  As expected, curves for the different bonds split at $j=\beta$ and $j=1$.}
\label{f:num2}
\end{center}
\end{figure}

In figure~\ref{f:num2} 
we show numerical results for the current fluctuations across the three different bonds in our toy system.  The results are in full agreement with the analytical expressions and graphically illustrate the spatial inhomogeneity of the fluctuations for large currents.

It is obvious that to study current fluctuations in a TAZRP with $L$ sites ($L+1$ bonds) requires an $L+1$ particle TASEP and hence the evaluation of an $(L+1) \times (L+1)$ determinant.  This becomes increasingly tedious for large $L$.  However, it is clear that each bulk particle with hopping rate one gives a factor $(1-z)^{-1}$ in the $D_0$ integrals of the determinant and hence that a saddle-point expansion for large times with careful treatment of the poles (as above) will lead to equations~(\ref{e:ql}) and~(\ref{e:qL}).

We note that the exclusion process picture gives further qualitative explanation for the behaviour of the current fluctuations in the different regimes.  One sees that for $j<1$ the bulk exclusion particles all pile up behind the leading particle, whereas for $j>1$ they move independently.  In other words, since particles cannot overtake, a bulk particle (with unit hopping rate) always has the same average current as the first particle if the latter is moving with speed less than one.  On the other hand, for a particle to move with current greater than unity requires all the particles in front of it to be moving \emph{independently} with $j>1$.  Similarly, the last particle in the TASEP is located immediately behind the penultimate particle for $j<\beta$ but moves independently for $j>\beta$.

\subsection{Conjecture for disordered TAZRP}
\label{ss:gencon}

Noting that the determinant solution~(\ref{e:detsol}) is valid for any choice of $v_l$ motivates us to consider the
behaviour of current fluctuations in the totally asymmetric zero-range process with bond-disordered rates.  
We consider an $L$ site TAZRP (again with $w_n=1$) where a particle moves from site $l$ to site $l+1$ with rate $r_l$.  The rate of hopping from the lefthand reservoir site into the system is denoted by $r_0$.  The pure case considered above corresponds to the choice $r_0=\alpha$, $r_L=\beta$ and $r_1, r_2, \dots, r_{L-1} =1$.   This problem can of course be represented as an $(L+1)$-particle exclusion process with hopping rates $v_l=r_{l-1}$.

One can easily show that the fugacities for the stationary state of the disordered TAZRP model are given by
\begin{eqnarray}
z_l = \frac{r_0}{r_l} \label{e:disfug} \\
\tilde{z}_l = 1,  \label{e:dislfug}
\end{eqnarray}
and thus the condition for a well-defined stationary state is $r_0 < r_l$ for all $l$.  In fact, the general conjecture to be presented below also describes current fluctuations in the case where this condition is violated and condensation occurs on one or more sites.

In our analysis of the current fluctuations we are interested in the product state with fugacites obtained from~(\ref{e:disfug}) by the substitution $r_0 \rightarrow r_0 \rme^{-\lambda}$.  If this is to be normalizable on site $l$ then we clearly require $r_0 \rme^{-\lambda} < r_l$ corresponding to the current condition $j<r_l$.  Since, in the totally asymmetric model, the current across a given bond is only dependent on the occupation of sites to the left it is clear that an important limiting r\^ole is played by the smallest hopping rate leftward from the bond under consideration.  Hence, for convenience, we define
\begin{equation}
r_l^*=\min_{k\leq l} \{r_k\}.
\end{equation}

Consideration of the form of the determinant solution~(\ref{e:detsol}) then leads to the following generalization of our pure results
\begin{equation}
\hat{e}_l(j)=
\cases{
r^*_l-j+j\ln\frac{j}{r^*_l} & $j < r^*_l$ \\ 
\sum_{k=0}^l \left( r_k-j+j\ln\frac{j}{r_k} \right) \Theta(j - r_k) & $j \geq r^*_l$ 
}
\end{equation}
where $\Theta(x)$ is the Heaviside step function.  Note, in particular, that the limiting behaviour of the current fluctuations is independent of the ordering of hopping rates leftward from the bond in question.  In fact it follows from~(\ref{e:detsol}) that, for an empty initial condition, this is also true for all finite times.

In other words, for currents less than the minimum hopping rate $r^*_l$ the current fluctuations are always controlled by the ``slowest bond''; for higher currents there are ``instantaneous condensates'' on all sites with exit hopping rate $r_l<j$.

\section{Discussion}
\label{s:dis}

In this paper we have considered the large-time limit of the current fluctuations in the zero-range process with open boundaries and some fixed initial configuration.  
For arbitrary boundary parameters and hopping rates we obtained a general result [equation~(\ref{e:elamb})] for the integrated current distribution and discussed the regime of its validity. 
We then provided explicit results for large current deviations in the totally asymmetric zero-range process (with hopping rate $w_n=1$) and showed how these can also be obtained via the application of new Bethe ansatz results for the totally asymmetric simple exclusion process with particle-dependent hopping rates.  Significantly, for boundary parameters giving a well-defined stationary state (i.e., no condensation), we found that for low currents the fluctuations are the same across all bonds but for higher currents they are spatially inhomogeneous.  We interpreted this change in behaviour as resulting from the presence of ``instantaneous condensates''---the \emph{temporary} build-up of particles on some site(s).   It is thus expected to be a generic feature for the TAZRP with other choices of $w_n$ for which the partition function $Z$ [see equation~(\ref{e:gc})] has a finite radius of convergence.  

We also expect to see similar behaviour for the partially asymmetric zero-range process (PAZRP) with the range of validity for the general result~(\ref{e:elamb}) dependent on the particular choice of $w_n$.  For currents larger than some critical current we again expect a change in character of the current fluctuations resulting from the appearance of ``instantaneous condensates''.  The PAZRP with $w_n=1$ can, of course, be mapped to the partially asymmetric exclusion process (PASEP); in principle, it should be possible to solve this mapped system by Bethe ansatz (at least for a small number of particles corresponding to small PAZRP system size, cf.\ \cite{Schutz97c} for particle-independent hopping rates) but a determinant form is not available.  Another consequence of the bidirectional movement of particles is that to determine fluctuations across the $l$th PAZRP bond we would need to solve a PASEP with $L+1$ particles rather than just $l+1$.

Finally, we remark that although our results can be applied to large systems by taking the thermodynamic limit $L \rightarrow \infty$, this limit does not necessarily commute with the long-time limit $t \rightarrow \infty$.  Our analysis thus has nothing to say about current fluctuations in a genuinely infinite system (taking the limit $L \rightarrow \infty$ first, followed by $t \rightarrow \infty$).  


\ack

AR is grateful for financial support from the Deutsche Forschungsgemeinschaft.  We thank Joel Lebowitz for useful discussions about the behaviour of current fluctuations in the steady state (different to the situation of fixed initial configuration that we consider here).

\section*{References}

\bibliographystyle{phaip}
\bibliography{/usr/users/iff_th2/rharris/allref}

\end{document}